\documentclass[journal]{IEEEtranTIE}
\usepackage{graphicx}
\usepackage{cite}
\usepackage{picinpar}
\usepackage{amsmath}
\usepackage{url}
\usepackage{flushend}
\usepackage[latin1]{inputenc}
\usepackage{colortbl}
\usepackage{soul}
\usepackage{multirow}
\usepackage{pifont}
\usepackage{color}
\usepackage{alltt}
\usepackage[hidelinks]{hyperref}
\usepackage{enumerate}
\usepackage{siunitx}
\usepackage{epstopdf}
\usepackage{pbox}
\usepackage[siunitx, RPvoltages]{circuitikz}
\usepackage{tikz}
\usepackage{amsmath}
\usepackage{amsfonts}
\usepackage{booktabs}
\usepackage{xcolor}
\usepackage{siunitx}
\usepackage{multicol}
\usepackage{algorithm2e}
\usepackage{microtype}
\usepackage{tcolorbox}

\tikzstyle{Unit}=[thick, draw=black, fill=black!20, circle, minimum size=20]
\usetikzlibrary{shapes.geometric}
\usetikzlibrary{positioning}

\DeclareMathOperator*{\argmin}{arg\,min}

\newif\ifblueR
\blueRfalse  

\ifblueR
    \newcommand{\highlightR}[1]{{\color{blue}#1}}
\else
    \newcommand{\highlightR}[1]{{\color{black}#1}}
\fi

\IEEEpubid{\parbox{\textwidth}{\footnotesize
\copyright~2024 IEEE.\ Personal use of this material is permitted.\ Permission from IEEE must be obtained for all other uses, in any current or future media, including reprinting/republishing this material for advertising or promotional purposes, creating new collective works, for resale or redistribution to servers or lists, or reuse of any copyrighted component of this work in other works.\\
Published in IEEE Transactions on Industrial Electronics.\ DOI: 10.1109/TIE.2024.3447734}}

\begin{document}
\title{Enhanced Imitation Learning of Model Predictive Control through Importance Weighting}

\author{
	\vskip 1em
	
	Pere~Izquierdo~G\'{o}mez, \emph{Student Member,~IEEE},
	Miguel~E.~L\'{o}pez Gajardo, \emph{Student Member,~IEEE},
	Nenad~Mijatovic, \emph{Senior Member,~IEEE},
	and Tomislav~Dragi\v{c}evi\'{c}, \emph{Senior Member,~IEEE}

	\thanks{
	
	The authors are with the Department of Wind and Energy Systems, Technical University of Denmark, Kongens Lyngby 2800, Denmark (email: \{pizgo, melga, nemi, tomdr\}@dtu.dk).

	}
}

\maketitle
 
\begin{abstract}
The approximation of model predictive control (MPC) algorithms with neural network models has received significant attention in the power electronics research community, as a valuable tool to enable the real-time implementation of computationally expensive methods.
Training machine learning models to imitate MPC algorithms typically follows a traditional supervised learning flow, with a static training data set generated through either uniform sampling or simulation.
Training with uniformly distributed data allows models to maintain consistent performance throughout the entire operating space, but very small models may fail to achieve satisfactory results in the regions corresponding to expected operation.
Conversely, simulation data can be used in training to obtain models that can accurately track certain trajectories but fail to achieve adequate performance in underrepresented regions.
The present article proposes a method to combine the advantages of both methods, by training models on sampled data with importance weights obtained through the kernel density estimation of experimental data.
This allows the training process to target specific regions of the operating space, and results in measurable performance improvements.
The method is demonstrated through the imitation of a moving discretized control set MPC algorithm for a dual active bridge converter, and is validated through experimental results.
\end{abstract}

\IEEEpubidadjcol


\markboth{IEEE TRANSACTIONS ON INDUSTRIAL ELECTRONICS}%
{}

\definecolor{limegreen}{rgb}{0.2, 0.8, 0.2}
\definecolor{forestgreen}{rgb}{0.13, 0.55, 0.13}
\definecolor{greenhtml}{rgb}{0.0, 0.5, 0.0}

\section{Introduction}

\IEEEPARstart{M}{odel} predictive control (MPC) has found success in a wide range of power electronics applications, owing to its ability to intuitively regulate multivariate and nonlinear systems with potentially complex control objectives and constraints~\cite{Vazquez2014}.
Despite its benefits, MPC carries a larger computational burden than many other control strategies.
This requirement can come into conflict with the high sampling frequencies encountered in many power electronic systems, commonly in the range of tens to hundreds of kilohertz, as in e.g.~\cite{Vazquez2017}.

An approach to reduce the computational burden of MPC is to approximate a particular algorithm with a neural network model, in a process referred to as \textit{imitation learning}.
This solution has recently received ample attention in the power electronics research community, having been employed on numerous converter topologies and explored in works including~\cite{Novak2021, Wang2021a, Wang2021b, Zafra2024, Simonetti2023, Mohamed2019, Wang2022, Liu2023}.
In MPC imitation, predetermined sets of inputs comprising external measurements and references are generated and run through the target MPC algorithm to result in input-control output pairs for supervised learning.
As MPC algorithms can generate a solution for every possible set of inputs, their input space can be probed arbitrarily, allowing for the generation of large quantities of training data.
The machine learning models employed in imitation learning tend to be relatively small, as the goal of the imitation process is to minimize the computational burden of the control system.
These characteristics stand in contrast to most applications of machine learning, where training data cannot be arbitrarily generated and model inference time is not as critical.
Specifically, the most commonly employed models in MPC imitation are feedforward fully-connected artificial neural networks (ANNs), with one~\cite{Novak2021, Wang2021a, Wang2021b, Simonetti2023, Mohamed2019, Wang2022, Liu2023} or two~\cite{Zafra2024} hidden layers.
Training data sets are typically generated by sampling the input space in a grid~\cite{Novak2021, Wang2021a, Wang2021b} or through dynamic simulation~\cite{Zafra2024, Simonetti2023, Mohamed2019, Wang2022}, with each method carrying certain advantages and disadvantages.
On the one hand, uniform grid sampling ensures that the training data set covers the entire operating space, but by evenly minimizing its prediction error, the resulting model may generate less accurate predictions in the regions most relevant to the particular application.
On the other hand, data generation through dynamic simulation requires meticulous design, as the resulting model can target specific regions of the input space but fail to generalize to those not contained in the simulated trajectories; to guarantee adequate performance beyond the training data, the design of experiments must sufficiently cover the potentially high-dimensional input space.

\IEEEpubidadjcol
\IEEEpubidadjcol

This article proposes a method to bridge the gap between sampling and simulation-based data generation, combining their advantages.
The method is based on importance-weighted training, where each training sample is assigned a weight proportional to its relevance, and the ANN model is trained accordingly.
To the best knowledge of the authors, this article presents the first application of importance-weighted training for the imitation of model predictive control.
Importance weighting (IW) has recently been successfully applied to neural network training problems in the context of reinforcement learning, where it has been used to increase learning speed by assigning greater weights to information-rich episodes~\cite{Wang2020}.
IW has also found success in online learning problems, where it has been used to mitigate the negative effects of concept drift~\cite{Fedeli2023}.

A procedure to obtain importance weights for a given imitation task is presented, based on empirical probability density estimation of either simulation or experimental data.
Combining the importance weights with uniformly sampled inputs allows for a tunable balance between targeting specific regions of the input space and ensuring that the model generalizes beyond them.
Through the use of the proposed method, model performance in these regions can be maintained even for very small model sizes.

As an illustrative example, consider an MPC algorithm where the output current and voltage measurements of a power converter are used as inputs to the model, and the load is known to be resistive.
For instance, for a two-quadrant power converter allowing for bidirectional current, generating training data through sampling within the limits of operation results in many training examples that would rarely be encountered, with a positive voltage and a negative current.
Through the use of importance-weighted training, the model can be made to target the first quadrant by assigning greater weights to samples inside of it, while still employing information from the second quadrant to ensure adequate generalization and robustness.

\highlightR{The main research contribution of this article is a novel methodology to incorporate application-specific knowledge to the training process of ANN-based imitators of control algorithms.
The proposed method is based on probability density estimation to assign greater importance during training to samples in the most relevant regions of the input space, such as the most frequently encountered operating setpoints.
In practice, this technique may enable the training of computationally lighter models that can match the performance of larger models within the targeted regions, allowing for real-time deployment in strictly-constrained computational environments or at higher control frequencies.
}

The proposed method is demonstrated through the imitation of a moving discretized control set MPC (MDCS-MPC) algorithm for a dual active bridge (DAB) power converter, including an experimental validation of the resulting ANN controller.
Section \ref{sec:methodology} presents the proposed methodology, summarizing the MDCS-MPC control algorithm, the model training process, and the procedure to compute the importance weights.
Section \ref{sec:results} presents both the training and the experimental verification results, and Section \ref{sec:conclusion} summarizes the main conclusions of the study.

\section{Methodology}\label{sec:methodology}
\highlightR{This section presents the dynamical model of the DAB converter,} the formulation of the MPC algorithm to be imitated, the importance-weighted training method, and a procedure to obtain application-specific importance weights.

\begin{figure}
    \centering
    \ctikzset{sources/scale=0.7}
\ctikzset{diodes/scale=0.4}
\ctikzset{inductors/scale=0.5}
\ctikzset{capacitors/scale=0.5}
\ctikzset{transistors/scale=0.5}
\ctikzset{bipoles/border margin=1} 
\ctikzset{nodes width/.initial=.02}
\ctikzset{current arrow scale=24}

\begin{circuitikz}[american]
    \draw (0, 3) node[ocirc]{} to [short, i=$i_{dc1}$, -*] (0.75, 3) to [capacitor, l={$C_1$}, v=$v_{dc1}$] (0.75, 0) to (0, 0) node[ocirc]{};

    \draw (1.5, 2.5) node[nigfete, bodydiode](Sp1){\footnotesize{$S_{p1}$}};
    \draw (1.5, 0.5) node[nigfete, bodydiode](Sp2){\footnotesize{$S_{p2}$}};
    \draw (2.75, 2.5) node[nigfete, bodydiode](Sp3){\footnotesize{$S_{p3}$}};
    \draw (2.75, 0.5) node[nigfete, bodydiode](Sp4){\footnotesize{$S_{p4}$}};

    \draw (0.75, 3) to [short, -*, i=$i_1$] (1.5, 3) to (Sp1.D);
    \draw (Sp1.S) to (Sp2.D);
    \draw (Sp2.S) to [short, -*] (1.5, 0) to [short, -*] (0.75, 0);
    \draw (1.5, 3) to (2.75, 3) to (Sp3.D);
    \draw (Sp3.S) to (Sp4.D);
    \draw (Sp4.S) to (2.75, 0) to (1.5, 0);

    \draw (4.25, 1.5) node[transformer core](T){\footnotesize$n:1$};
    \draw (T.A1) to [inductor, l=$L$, mirror] ++(-0.7, 0) to [short, -*] ++(-1.525, 0);
    \draw (T.A2) to [short, -*] ++(-0.975, 0);
    \draw (T.B1) to [short, -*] ++(0.225, 0);
    \draw (T.B2) to [short, -*] ++(1.475, 0);
    
    \draw (5, 2.5) node[nigfete, bodydiode](Ss1){\footnotesize{$S_{s1}$}};
    \draw (5, 0.5) node[nigfete, bodydiode](Ss2){\footnotesize{$S_{s2}$}};
    \draw (6.25, 2.5) node[nigfete, bodydiode](Ss3){\footnotesize{$S_{s3}$}};
    \draw (6.25, 0.5) node[nigfete, bodydiode](Ss4){\footnotesize{$S_{s4}$}};

    \draw (Ss1.D) to [short, -] (5, 3) to [short, -*] (6.25, 3) to (Ss3.D);
    \draw (Ss1.S) to (Ss2.D);
    \draw (Ss2.S) to [short, -] (5, 0) to [short, -*] (6.25, 0) to (Ss4.S);
    \draw (Ss3.S) to (Ss4.D);

    \draw (6.25, 3) to [short, -*, i=$i_2$] (7.5, 3) to [short, i=$i_{dc2}$] (8.25, 3) node[ocirc]{};
    \draw (7.5, 3) to [capacitor, l={$C_2$}, v=$v_{dc2}$] (7.5, 0);
    \draw (6.25, 0) to [short, -*] (7.5, 0) to (8.25, 0) node[ocirc]{};
\end{circuitikz}
    \vspace{-0.7cm}
    \caption{Circuit diagram of the dual active bridge (DAB) converter.}\label{fig:dab_circuit_diagram}
\end{figure}
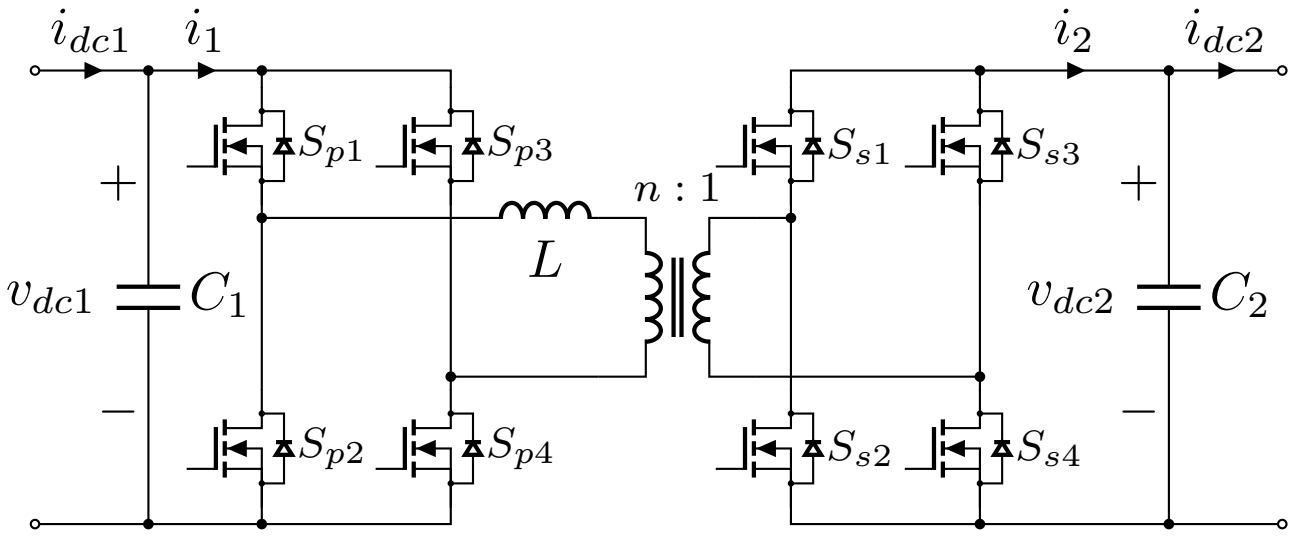

\highlightR{\subsection{Modeling of the DAB converter}
A circuit diagram of the DAB power converter under study is shown in Fig.~\ref{fig:dab_circuit_diagram}.
It is a bidirectional dc-dc converter topology which consists of two full-bridges coupled through a power transfer inductance $L$ and a high-frequency transformer with an $n:1$ turns ratio.
Considering input and output capacitors $C1$ and $C2$, respectively, the voltages $v_{dc1}$ and $v_{dc2}$ on each side of the converter can be modeled as
\begin{equation}
    C_1 \frac{\mathrm{d} v_{dc1}}{\mathrm{d} t} = i_{dc1} - i_1, \quad C_2 \frac{\mathrm{d} v_{dc2}}{\mathrm{d} t} = i_2 - i_{dc2}.
    \label{eq:dab_continuous}
\end{equation}

Using the forward Euler method, the output voltage equation in \eqref{eq:dab_continuous} can be discretized to obtain
\begin{equation}
    v_{dc2}^{k} = \frac{1}{C_2 f_c} \left( i_2^{k-1} - i_{dc2}^{k-1} \right) + v_{dc2}^{k-1}
    \label{eq:dab_discrete}
\end{equation}
}

\subsection{The MDCS-MPC Algorithm}

The ANN is trained to imitate an MDCS-MPC algorithm for a DAB converter, which has been presented for output voltage~\cite{Chen2020, Chen2020a} and output current~\cite{Chen2020b} control.
The MDCS-MPC is designed for output voltage ($v_{dc2}$) control, and is based on~\cite{Lopez2023, Chen2020}.
This algorithm employs a dynamical model of the converter to obtain an optimal current actuation $\bar{i}_{2}$, averaged over its sampling period.
$\bar{i}_{2}$ is then used in Triple-Phase Shift (TPS) modulation to synthesize the switching signals, following the method presented in~\cite{Shao2019}.
In order to avoid steady-state error due to model inaccuracy, a current compensation term based on the current actuation error is introduced to the model. 
The reader is referred to \cite{Lopez2023, Chen2020} for a complete description of this control method.
This term is defined as
\begin{equation}
    i_{comp}^{k-1} = \text{LPF}\left(\hat{i}_2^{k-1} - \bar{i}_2^{k-1} \right),
\end{equation}
where LPF stands for a low-pass filter, $k-1$ is the present measurement time step, and $\hat{i}_2^{k-1}$ is the estimated actuation \highlightR{derived from \eqref{eq:dab_discrete} as}
\begin{equation}
    \hat{i}_2^{k-1} = C_2 f_c \left( v_{dc2}^{k-1} - v_{dc2}^{k-2} \right) + i_{dc2}^{k-2}.
\label{eq:i2hat}
\end{equation}

In \eqref{eq:i2hat}, $f_c$ is the control frequency, while $v_{dc2}$ and $i_{dc2}$ are obtained from measurement.
Following~\cite{Cortes2012}, delay compensation is implemented as a one-step prediction using the system model, the measurements, and the actuation at time $k-1$, as
\begin{equation}
    v_{dc2}^k = \frac{1}{C_2 f_c} \left( \bar{i}_2^{k-1} + i_{comp}^{k-1} - i_{dc2}^{k-1} \right) + v_{dc2}^{k-1}.
\end{equation}

The MDCS-MPC algorithm employs two-step predictions, the first of which is given by
\begin{equation}
    v_{dc2}^{k+1} = \frac{1}{C_2 f_c} \left( \bar{i}_2^{k} + i_{comp}^{k-1} - i_{dc2}^{k} \right) + v_{dc2}^{k},
\end{equation}
making use of the assumption that $i_{comp}^{k-1}$ remains approximately constant in consecutive periods. Additionally, using the approximation $i_{dc2}^{k+1} \approx i_{dc2}^{k}$, the second prediction can be expressed as
\begin{equation}
    v_{dc2}^{k+2} = \frac{1}{C_2 f_c} \left( \bar{i}_2^{k+1} + i_{comp}^{k-1} - i_{dc2}^{k} \right) + v_{dc2}^{k+1}.
\end{equation}

The optimal actuation can then be found by minimizing the cost function, as
\begin{equation}
    \bar{i}_2^{k} = \argmin_{\bar{i}_2^{k}, \bar{i}_2^{k+1}} \underbrace{\left( v_{dc2}^{*} - v_{dc2}^{k+2} \right)^2 + \lambda \left( v_{dc2}^k - v_{dc2}^{k+2} \right)^2}_{\text{Cost function}},
    \label{eq:cost_function_mpc}
\end{equation}
where $v_{dc2}^{*}$ is the output voltage reference for the instant $k+2$ and $\lambda \geq 0$ is a tunable weighting factor.
The first term of the objective function penalizes tracking error, while the second serves to enforce output voltage convergence \highlightR{by penalizing controller trajectories that would lead to oscillatory behavior, as shown in~\cite{Chen2020}}.
In order to solve the optimization problem at the required control frequency, MDCS-MPC discretizes the two variables $\bar{i}_2^{k}$ and $\bar{i}_2^{k+1}$ according to
\begin{equation}
    \bar{i}_{2, c}^k = \bar{i}_2^{k-1} + \delta_{adp} \left( c - \frac{\mu - 1}{2} \right) \; \forall c \in \{0, 1, \dots, \mu -1\},
    \label{eq:candidate_currents}
\end{equation}
where $\delta_{adp} = \delta_i \left(1 + \gamma i_{\delta}^2 \right)$, $\mu$ is the number of discretization steps, \highlightR{$c$ is an index variable corresponding to each actuation candidate}, $\delta_i$ and $\gamma$ are two gain factors, and $i_{\delta}$ is given by
\begin{equation}
    i_{\delta} =
    \begin{cases}
        I_{max} \left| \Delta \right| / \Delta_{max}, & \text{if} \left| \Delta \right| < \Delta_{max}, \\
        I_{max}, & \text{otherwise}.
    \end{cases}
\end{equation}

$\Delta_{max}$ and $I_{max}$ are defined as constant limits, while $\Delta = v_{dc2}^* - v_{dc2}^{k-1}$. 
The average output current limit $\bar{I}_{2, max}$ over the sampling period is adjusted according to the input voltage, as
\begin{equation}
    \bar{I}_{2, max} = \frac{n v_{dc1}}{8 f_{sw} L},
\end{equation}
where $n$ is the transformer turns ratio and $f_{sw}$ is the switching frequency, i.e.\ operating frequency of the modulator.
\highlightR{If one or more candidate currents $\bar{i}_{2, c}^k$ as obtained from \eqref{eq:candidate_currents} exceed $\bar{I}_{2, max}$, they are replaced with this limit ahead of the evaluation of the model and the cost function, ensuring that the maximum power transfer and the converter actuation capability are respected.}

\highlightR{The optimization of the cost function \eqref{eq:cost_function_mpc} is performed through exhaustive search, requiring $\mu^2$ evaluations of the system states and objectives.}

\subsection{Importance-Weighted Training}
In the existing literature, the training data for MPC imitators are commonly generated through sampling the input space in a grid~\cite{Novak2021, Wang2021a, Wang2021b} or based on dynamic simulation data~\cite{Zafra2024, Simonetti2023, Mohamed2019, Wang2022}.
In this manner, the training data are defined as a set of input-output pairs $\mathcal{D} = \{\left(\mathbf{x}_1, \mathbf{y}_1\right), \dots, \left(\mathbf{x}_N, \mathbf{y}_N\right)\}$, with inputs obtained through sampling and outputs found by evaluating the MPC algorithm on the set of inputs.
The supervised training of neural networks can be understood from the perspective of empirical risk minimization~\cite{Goodfellow2016}, which defines the risk function $\mathcal{L}$ as the expectation of a loss function $\ell$, given by
\begin{equation}
    \mathcal{L}(\theta) =  \mathbb{E}_{(\mathbf{x}, \mathbf{y}) \sim p_{train}} \left[ \ell \left( f \left(\mathbf{x}; \theta \right), \mathbf{y} \right) \right],
\end{equation}
where $f(\cdot, \theta)$ is a neural network model with parameters $\theta$, and $p_{train}$ is the underlying probability distribution of the training data $\mathcal{D}$.
Since $p_{train}$ is defined by either a sampling method or a limited collection of dynamic simulations, it does not directly correspond to the probability distribution observed during field operation, $p_{test}$.
To compensate for this mismatch while avoiding sampling from the unknown distribution $p_{test}$, importance sampling methods redefine the risk as~\cite{Fang2020}:
\begin{align}
    \mathcal{L}'(\theta) &=  \mathbb{E}_{(\mathbf{x}, \mathbf{y}) \sim p_{test}} \left[ \ell \left( f \left(\mathbf{x}; \theta \right), \mathbf{y} \right) \right] \\
    & =  \mathbb{E}_{(\mathbf{x}, \mathbf{y}) \sim p_{train}} \left[ \frac{p_{test}(\mathbf{x}, \mathbf{y})}{p_{train}(\mathbf{x}, \mathbf{y})} \ell \left( f \left(\mathbf{x}; \theta \right), \mathbf{y} \right) \right]. \label{eq:empirical_risk}
\end{align}

Importance weights can then be defined as proportional to the likelihood ratio, i.e.\ 
\begin{equation}
    w(\mathbf{x}, \mathbf{y}) \propto \frac{p_{test}(\mathbf{x}, \mathbf{y})}{p_{train}(\mathbf{x}, \mathbf{y})}, \label{eq:importance_weights_ratio}
\end{equation}
which in the case where $p_{train}$ is a uniform distribution, become
\begin{equation}
    w(\mathbf{x}, \mathbf{y}) \propto p_{test}(\mathbf{x}, \mathbf{y}). \label{eq:importance_weights}
\end{equation}

Since the risk defined in \eqref{eq:empirical_risk} is minimized during the training process, proportionality will not affect the optimal $\theta$.
The proposed method then defines a set of importance weights $\mathcal{W} = \{w_1, \dots, w_N\}$, where a weight $w_i \in \mathbb{R}_{\geq 0}$ is assigned to every element of $\mathcal{D}$.

The ANN model is trained with \highlightR{mini-batch} stochastic gradient descent (SGD) or one of its variants, such as SGD with momentum or Adam~\cite{Kingma2014}, which employ data batches $\mathcal{B}$ defined as randomly selected subsets of the joint set $\mathcal{D} \cup \mathcal{W} = \{\left(\mathbf{x}_1, \mathbf{y}_1, w_1 \right), \dots, \left(\mathbf{x}_N, \mathbf{y}_N, w_N \right)\}$.
\highlightR{Batch training reduces the variance of the loss gradient estimates compared to single-sample updates, resulting in more stable and efficient training processes.}
The importance-weighted mean squared error (MSE) on each batch, which follows from \eqref{eq:empirical_risk}, is defined as
\begin{equation}
    \mathcal{L}\left(\mathcal{B}, f, \theta \right) = \frac{1}{|\mathcal{B}|} \sum_{i=1}^{|\mathcal{B}|} w_i \left( \mathbf{y}_i - f \left( \mathbf{x}_i, \theta \right) \right)^2,
    \label{eq:mse_loss}
\end{equation}
where $|\mathcal{B}|$ represents the cardinality or total number of elements in the batch $\mathcal{B}$.
MSE loss is commonly used only in regression problems, but importance weighting can be included in any loss function, such as cross-entropy for classification problems.
Using SGD without momentum, the model parameters are updated as 
\begin{equation}
    \theta_{t+1} \leftarrow \theta_t - \eta \nabla_\theta \mathcal{L} \left( \mathcal{B}_t, f, \theta \right) \rvert_{\theta=\theta_t},
    \label{eq:sgd_update}
\end{equation}
where $\eta > 0$ is a tunable learning rate, and the gradients $\nabla_\theta \mathcal{L} \left( \mathcal{B}_t, f, \theta \right) \rvert_{\theta=\theta_t}$ are obtained through backpropagation and evaluated at the current values of $\theta$.
Without IW, i.e. setting every importance weight to 1, each input-output pair has the same impact on parameter updates during the training process.

Importance weighting has been shown to have a limited effect on overparameterized models past the initial training stages, as sufficiently large models are able to minimize the contribution to the total loss of each individual sample regardless of its scaling~\cite{Byrd2018, Xu2021}.
However, MPC imitation problems typically involve large amounts of training data and small-sized models, meaning that overparameterization will hardly ever be encountered.

\subsection{Density Estimation}
The key to the effectiveness of the weighted training is to adequately define the set of importance weights $\mathcal{W}$.
Following \eqref{eq:importance_weights_ratio}, the weights should be defined in terms of the estimated probability density of the target distribution, $\hat{p}_{test}$.
Intuitively, the weights should steer the training process towards lowering the prediction error in the most relevant regions of the input space, at the cost of a higher error outside of them.
Conventional distribution fitting can be employed to define $\hat{p}_{test}$, but the design of the density function in a relatively high-dimensional input space may pose a significant challenge.

The proposed method instead relies on kernel density estimation (KDE), which can estimate an empirical probability density function corresponding to any set of data with minimal assumptions on the shape of the distribution.
To this end, KDE employs a kernel function $K$, which assigns a density estimate $\hat{p}_{test}\left( \mathbf{x} \right) \geq 0$ to any arbitrary point $\mathbf{x} \in \mathbb{R}^d$ in the $d$-dimensional input space, according to 
\begin{equation}
    \hat{p}_{test}\left( \mathbf{x}\right) = \frac{1}{mh} \sum_{i = 1}^m K \left( \frac{\mathbf{x}- \tilde{\mathbf{x}}_{i}}{h} \right),
\end{equation}
where $h > 0$ is a tunable smoothing parameter commonly called bandwidth, \highlightR{and $m$ is the number of samples in $\mathcal{M}$}.
The KDE generates predictions based on a set of simulation or experimental measurements $\mathcal{M} = \{ \tilde{\mathbf{x}}_{i}, \dots, \tilde{\mathbf{x}}_{m} \}$.
There are many possible choices of kernel function; due to its interpretability and smoothness, the presented method employs a Gaussian kernel~\cite{Silverman2018}, defined as
\begin{equation}
    K(\mathbf{u}) = \frac{1}{\sqrt{2\pi}} \exp{\left(-\frac{1}{2} \| \mathbf{u} \|^2 \right)}.
\end{equation}

\highlightR{Kernel functions with bounded support, such as the Epanechnikov or the uniform kernel, may be preferable in cases where $\mathcal{M}$ contains a large number of samples.
The implementation of this type of kernel functions can be significantly more computationally efficient than that of unbounded support kernels like the Gaussian, thus reducing the KDE evaluation time necessary to obtain $\mathcal{W}$.}
The importance weights in $\mathcal{W}$ are then obtained for every sample in $\mathcal{D}$, as
\begin{equation}
    w_i = \frac{\hat{p}_{test}\left( \mathbf{x}_{i} \right)}{\displaystyle \max_{1 \leq j \leq N} \hat{p}_{test} \left( \mathbf{x}_{j} \right)},
    \label{eq:iw_normalized}
\end{equation}
\highlightR{where $N$ is the number of samples in $\mathcal{D}$}.
Normalizing the KDE estimates by their maximum value over $\mathcal{D}$ ensures that the loss function can only be scaled down, thus minimizing the risk of the importance weights introducing instability in the training process.
Without this normalization step, and depending on employed kernel, the density estimates could reach values $\hat{p}_{test} \left( \mathbf{x}_{i} \right) \gg 1$, resulting in larger gradient update steps than intended.

In conclusion, the presented KDE-based method to determine the importance weights only introduces a single tunable parameter: the bandwidth $h$ of the kernel function, which for a Gaussian kernel can be interpreted as its standard deviation.
Higher values of $h$ will lead to higher importance weights away from the set of measurements, thus leading the model to learn all regions of the input space more evenly.
Alternatively, lower values of $h$ will lead to weights that assign higher penalties to the loss close to the set of measurements, thus leading to better model performance in these regions at the cost of increased error away from them.

\begin{figure}
    \centering
    \includegraphics[]{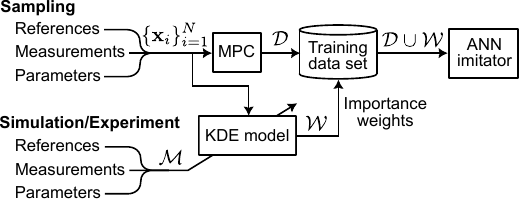}
    \vspace{-0.0cm}
    \caption{Block diagram of the training process with importance weighting.}\label{fig:block_diagram}
\end{figure}

The proposed method, including the process to obtain importance weights and the training process of the model, is detailed in Algorithm~\ref{alg:iw_training} and summarized in the diagram shown in Fig.~\ref{fig:block_diagram}.
\highlightR{The convergence criterion for the presented models is defined as a maximum number of epochs, which are defined as a complete iteration through $\mathcal{D}$.}

\RestyleAlgo{ruled}
\begin{algorithm}
\caption{Proposed importance-weighted training of MPC surrogate models.}\label{alg:iw_training}
\textit{\textbf{Require:}}\\
An initialized ANN $f: \mathcal{X} \rightarrow \mathcal{Y}$ with parameters $\theta_0$. \\
A training loss function $\ell: \mathcal{Y} \times \mathcal{Y} \rightarrow \mathbb{R}$. \\
A uniformly-sampled training set $\mathcal{D} = \{\left(\mathbf{x}_i, \mathbf{y}_i\right) \}_{i=1}^N$. \\
A simulation or experimental set $\mathcal{M} = \{ \tilde{\mathbf{x}}_{i} \}_{i=1}^m$. \\
A kernel function $K: \mathcal{X} \rightarrow \mathbb{R}$. \medskip

Define the set of importance weights $\mathcal{W} = \{w_i \}_{i=1}^N$ as
\begin{equation*}
    \hat{p}_{test}\left( \mathbf{x}_i\right) = \frac{1}{mh} \sum_{j = 1}^m K \left( \frac{\mathbf{x}_i - \tilde{\mathbf{x}}_{j}}{h} \right), \quad \eqref{eq:iw_normalized}. 
\end{equation*}

\While{not converged} {
Draw a random batch $\mathcal{B}_t$ from $\mathcal{D} \cup \mathcal{W}$. \\
Compute the importance-weighted MSE loss \eqref{eq:mse_loss}.\\
Using backpropagation and SGD, update $\theta$ \eqref{eq:sgd_update}.\\
$t \leftarrow t+1$
}
\end{algorithm}

\section{Results} \label{sec:results}
This section details the implementation-specific design choices in the importance-weighted training process and presents the corresponding experimental results.

\subsection{Model Training}
\begin{figure}
    \centering
    \begin{tikzpicture}[x=2.2cm,y=1.4cm]
  \tikzset{every node/.style={inner sep=0pt}}
    \def\inputnames{{"x_1", "x_2", "x_3", "x_4", "x_5", "x_6", "x_n"}}
    \def\yshift{0.1}
    \foreach \N [count=\lay,remember={\N as \Nprev (initially 0);}]
                 in {6,5,1}{ 
      \foreach \i [evaluate={\c=int(\i==\N)*int(\N>1)*int(\N==2); \y=\N/2-\i-\c*\yshift; \x=\lay; \prev=int(\lay-1);}]
                   in {1,...,\N}{ 
        \ifnum\i<\N
          \node[Unit, outer sep=0.6] (N\lay-\i) at (\x,\y) {$a^{(\prev)}_\i$};
        \fi
        \ifnum\i=\N
          \ifnum\N>1
            \ifnum\lay=1
              \node[Unit, outer sep=0.6] (N\lay-\i) at (\x,\y) {$a^{(\prev)}_{6}$};
            \fi
            \ifnum\lay>1
              \node[Unit, outer sep=0.6] (N\lay-\i) at (\x,\y) {$a^{(\prev)}_{n}$};
            \fi
          \fi
          \ifnum\N=1
            \node[Unit, outer sep=0.6] (N\lay-\i) at (\x,\y) {$a^{(\prev)}_1$};
          \fi
        \fi
        \ifnum\Nprev=0
          \ifnum\i<\N
            \draw[thick, -stealth] (\x-0.5, \y) to node[above]{} (N\lay-\i);
          \fi
          \ifnum\i=\N
            \draw[thick, -stealth] (\x-0.5, \y) to node[above]{} (N\lay-\i);
          \fi
        \fi

        \ifnum\N=1
            \draw[thick, -stealth] (N\lay-\i) to node[above=0.1]{$\bar{i}_2^{k}$} ++(0.5, 0);
        \fi
        
        \ifnum\Nprev>0
            \ifnum\i=\N
                \ifnum\N>1
                    \path (\x, \y) -- ++(0, 1.1+\yshift) node[midway, scale=1.4] {$\vdots$};
                \fi
            \fi
        \fi
        
        \ifnum\Nprev>0 
          \foreach \j in {1,...,\Nprev}{ 
            \ifnum\N>1 
              \draw[thick] (N\prev-\j) -- (N\lay-\i);
            \fi
            \ifnum\N=1 
              \ifnum\j<\Nprev
                \draw[thick] (N\prev-\j) -- (N\lay-\i)
                  node[pos=0.40]{\begin{tcolorbox}[colback=white,
                  colframe=white,
                  hbox, size=fbox,
                  arc=5mm, auto outer arc,
                 ]
                 $W^{(2)}_{\j,1}$
                \end{tcolorbox}};
              \fi
              \ifnum\j=\Nprev
                \draw[thick] (N\prev-\j) -- (N\lay-\i)
                node[pos=0.40]{\begin{tcolorbox}[colback=white,
                  colframe=white,
                  hbox, size=fbox,
                  arc=5mm, auto outer arc,
                 ]
                 $W^{(2)}_{n,1}$
                \end{tcolorbox}};
              \fi
            \fi
          }
        \fi
      }
    }

    \node at (N1-1)[above=0.5cm, align=center] {\small Input layer};
    \node at (N2-1)[above=0.5cm, align=center] {\small Hidden layer}; 
    \node at (N3-1)[above=0.5cm, align=center] {\small Output\\\small layer};
    
    \node at (N1-1)[shift={(N1-1.west)}, xshift=-0.35cm, yshift=0.25cm] {\small $v_{dc2}^{*}$};
    \node at (N1-2)[shift={(N1-2.west)}, xshift=-0.35cm, yshift=0.25cm] {\small $i_{dc2}^{k}$};
    \node at (N1-3)[shift={(N1-3.west)}, xshift=-0.35cm, yshift=0.25cm] {\small $v_{dc2}^{k}$};
    \node at (N1-4)[shift={(N1-4.west)}, xshift=-0.35cm, yshift=0.25cm] {\small $v_{dc1}^{k}$};
    \node at (N1-5)[shift={(N1-5.west)}, xshift=-0.35cm, yshift=0.25cm] {\small $\bar{i}_2^{k-1}$};
    \node at (N1-6)[shift={(N1-6.west)}, xshift=-0.35cm, yshift=0.25cm] {\small $i_{comp}^{k-1}$};
    
\end{tikzpicture}
    \highlightR{\caption{Structure of the fully-connected ANN model used as an imitator, with $n$ units in its single hidden layer.}
    \label{fig:feedforward_nn}}
\end{figure}
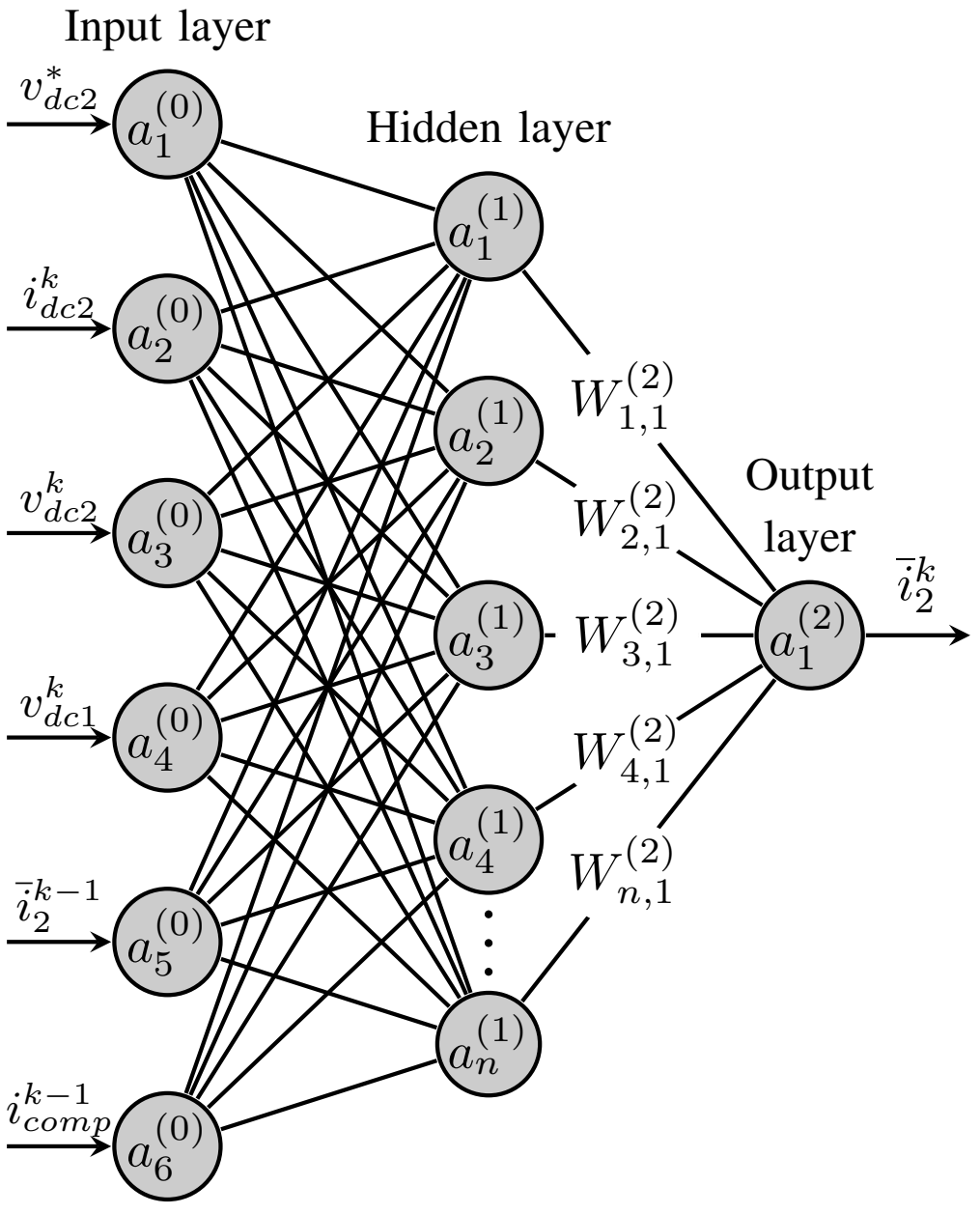

In a manner consistent with the relevant literature in MPC imitation within power electronics, the model selected to approximate the MDCS-MPC algorithm is a feedforward ANN with a single hidden layer.
\highlightR{Models based on this simple architecture can often be computationally light enough for their real-time evaluation at the high-frequencies required in power electronics applications.}
In particular, the model receives six inputs: $\{v_{dc2}^{*}, i_{dc2}^{k}, v_{dc2}^{k}, v_{dc1}^{k}, \bar{i}_2^{k-1}, i_{comp}^{k-1}\}$, and it outputs a scalar estimate of the reference current $\bar{i}_2^{k}$.
\highlightR{This model architecture is represented schematically in Fig.~\ref{fig:feedforward_nn}.}
The operating limits of each dimension, employed both in sampling and in data normalization, are collected in Table~\ref{tab:operating_limits}.

\begin{table}
    \vspace{-1em}
        \caption{Operating limits of each input variable}\label{tab:operating_limits}
        \centering
        \bgroup
        \def\arraystretch{1.3}
        \begin{tabular}{c l c c}\toprule
            Symbol & Description & Min. & Max. \\
            \midrule
            \rowcolor{gray!25} $v_{dc2}^{*}$ & Output voltage reference & \qty{90}{V} & \qty{240}{V} \\ 
            $i_{dc2}^{k}$ & Output current measurement & \qty{-12}{A} & \qty{12}{A} \\ 
            \rowcolor{gray!25} $v_{dc2}^{k}$ & Output voltage measurement & \qty{0}{V} & \qty{250}{V} \\ 
            $v_{dc1}^{k}$ & Input voltage measurement & \qty{0}{V} & \qty{330}{V} \\ 
            \rowcolor{gray!25} $\bar{i}_2^{k-1}$ & Previous controller output & \qty{-10}{A} & \qty{10}{A} \\ 
            $i_{comp}^{k-1}$ & Compensation current & \qty{-2}{A} & \qty{32}{A} \\
            \bottomrule
        \end{tabular}
        \egroup
    \end{table}

\begin{figure}
    \centering
    \includegraphics[]{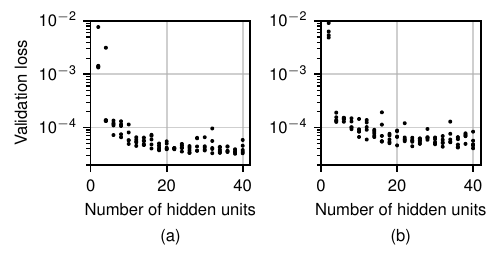}
    \vspace{-0.4cm}
    \highlightR{\caption{Final validation loss as a function of the number of hidden units, corresponding to (a) baseline and (b) importance-weighted models.}\label{fig:pareto_loss}}
\end{figure}

\highlightR{In order to select the number of units in the hidden layer, multiple models were trained with different hidden unit counts, ranging from 2 to 40, and employing the same parameters shown in Table~\ref{tab:operating_limits}.
The final validation loss corresponding to training five of each of these models is displayed in Fig.~\ref{fig:pareto_loss}.
The number of units in the hidden layer was therefore selected at 10, as the minimum value that allows for models that were able to experimentally control the DAB without crossing any operation or safety limits, and past which the gains in model accuracy start to become limited, as shown in Fig.~\ref{fig:pareto_loss}.}
All units in the hidden layer employed the rectified linear unit (ReLU) activation function.
\highlightR{The output layer contains a single unit with a linear activation. 
To ensure that the operating limits of the converter are not exceeded, the triple-phase shift modulator will effectively cap current references to their upper and lower limits, as described in~\cite{Lopez2023}.}
The model training process was implemented in PyTorch 2.0, running on Python 3.11.
\highlightR{The optimizer used to train all presented models was Adam, with an empirically selected fixed learning rate and its other parameters fixed at the default values suggested in \cite{Kingma2014}.}
\highlightR{The models' parameters were initialized following a Kaiming~\cite{He2015} uniform distribution, with biases set to 0 and weights $W$ obtained as
\begin{equation}
    W \sim \mathcal{U} \left(-\frac{1}{\sqrt{n_{l-1}}}, \frac{1}{\sqrt{n_{l-1}}} \right),
\end{equation}
where $n_{l-1}$ is the number of units in the previous layer.}

Training data were generated using uniform random sampling, allowing for \eqref{eq:importance_weights} to hold without additional considerations.
For this particular problem, uniform sampling was also found to result in a lower loss during training, both with and without importance weighting, than the more frequently used grid sampling, and a similar loss to the also common latin hypercube sampling~\cite{McKay1979}.
Additionally, random sampling generalizes better than grid sampling to input spaces with a higher dimensionality, as the number of samples obtained from a grid scale exponentially with the number of dimensions of the input space.
The sampled inputs were run through the MDCS-MPC algorithm to generate a total of one million training samples, a number above which the relatively small model would not improve its performance significantly.
An additional $10^5$ random samples were employed as the validation set for the training process.
All values were normalized with min-max scaling for both the computation of importance weights and for model training and inference.
\highlightR{No regularization was employed during training, as no overfitting was observed during training due to the large number of training samples relative to the size of the models.}

Importance weights were generated through a KDE model with a Gaussian kernel and a bandwidth of $h = 0.4$, which resulted in the best-performing model among a grid of values defined as $\{0.2, 0.4, 0.5, 0.6, 0.8\}$.
The KDE model was implemented with the scikit-learn library~\cite{scikit-learn} and fit to an experimental data set $\mathcal{M}$ obtained from a single experiment on the DAB setup with the MDCS-MPC algorithm.

\begin{figure}
    \centering
    \includegraphics[]{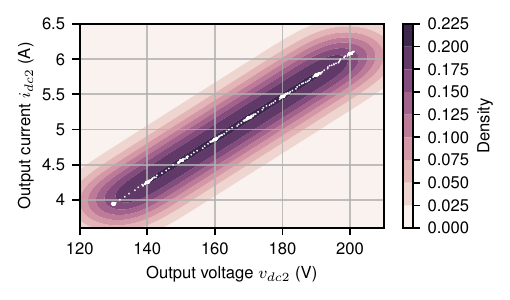}
    \vspace{-0.5cm}
    \caption{KDE of a set of output current and voltage measurements, based on 10.000 data points (in white).}\label{fig:lab_voltage_current_KDE}
\end{figure}

Fig.~\ref{fig:lab_voltage_current_KDE} presents an example illustration of the fitting of a KDE model on a set of output current $i_{dc2}$ and output voltage $v_{dc2}$ measurements obtained by applying the MDCS-MPC on the DAB converter with a resistive load rated at \qty{30.6}{\ohm}.
The figure serves only for illustrative purposes, as the target MDCS-MPC algorithm receives 6-dimensional inputs, meaning that the KDE also generates predictions on 6-dimensional feature space; the problem is then reduced to two dimensions for the sole purpose of visualization.
The KDE model is fit to the measurements, in white, with a bandwidth of $h = 0.4$, and the density predictions are graphed as contours.
The figure shows how, with a resistive load, there is a clear linear relationship between the two variables which the KDE model is able to capture.
In this case, the importance-weighted model would be able to minimize its prediction error in the regions close to the measurements, in a manner proportional to the kernel density estimation.
\highlightR{As shown by the results that follow, these observations are also applicable to input spaces with a higher dimensionality.}

To act as a baseline, equivalent ANN models were trained without importance weighting.
As an additional baseline, models were trained directly on $\mathcal{M}$, but these were unable to control the DAB within its defined operation limits and their results are therefore not reported.
This behavior suggests that models trained exclusively on simulation or experimental data may fail to learn the overall behavior of the MPC algorithm, leading to unpredictable performance.

\begin{table}
    \vspace{-1em}
        \caption{Model and training parameters}\label{tab:training_process}
        \centering
        \begin{tabular}{l c}\toprule
            Parameter & Value \\
            \midrule
            \rowcolor{gray!25} Optimizer & Adam \\
            Learning rate ($\eta$) & $10^{-4}$ \\
            \rowcolor{gray!25} Batch size ($|\mathcal{B}|$) & $8$ \\
            Number of model inputs & $6$ \\
            \rowcolor{gray!25} Number of hidden layers & $1$ \\
            Number of hidden units & $10$ \\
            \rowcolor{gray!25} Number of outputs & $1$ \\
            Size of the training set ($|\mathcal{D}|$) & $10^6$ \\
            \rowcolor{gray!25} Size of the experimental set ($|\mathcal{M}|$) & $10^5$ \\
            KDE bandwidth ($h$) & $0.4$ \\
            \rowcolor{gray!25} Number of training epochs & 100 \\
            \bottomrule
        \end{tabular}
    \end{table}

\begin{figure}
    \centering
    \includegraphics[trim=6 0 0 0, clip]{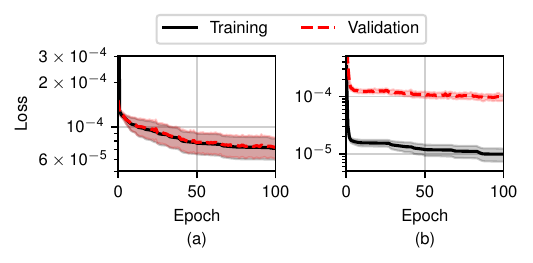}
    \vspace{-0.9cm}
    \highlightR{\caption{Loss curves corresponding to (a) the conventional and (b) the importance-weighted training processes, with 95\% confidence intervals obtained over 20 repetitions. Note that the training loss in (b) takes into consideration the importance weights, which results in a significant reduction compared to the training loss in (a). The validation loss, computed on a uniformly sampled data set and without importance weights, is expectedly higher for the IW training.}
    \label{fig:loss_curves}}
\end{figure}

The parameters corresponding to the training process are summarized in Table~\ref{tab:training_process}. 
Each stochastic data generation and training process was repeated multiple times to ensure consistency, with the results that follow corresponding to the best-performing model in each category.
\highlightR{Fig.~\ref{fig:loss_curves} shows the training and validation loss curves for both types of training, collected over 20 training processes with varying ANN parameter initializations and data sets.}

\subsection{Experimental Results}
\begin{figure}
    \centering
    \includegraphics[]{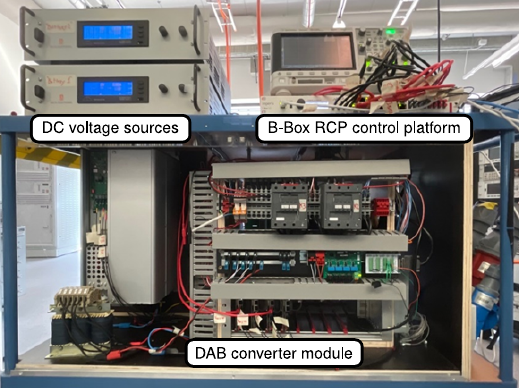}
    \vspace{-0.5cm}
    \caption{The DAB experimental setup employed in this study.}\label{fig:dab_setup}
\end{figure}

The presented control methods were validated on an experimental test bench, which comprises a Delta Elektronika dc source, four Imperix PEB8024 SiC MOSFET half-bridges, an array of 16 series-connected Bourns PQ2617BHA-4R7K inductors forming the power transfer inductance $L$, a Himag planar transformer, and a resistive load.
An Imperix B-Box rapid control prototyping (RCP) system was used as the control and monitoring platform, responsible for executing the control algorithms in real-time, sending the optical switching signals to the half bridges, and recording the measurements.
The employed experimental setup is shown in Fig.~\ref{fig:dab_setup}.

The most relevant parameters corresponding to the DAB experimental setup and the considered MDCS-MPC algorithm are reported in Table~\ref{tab:parameters}.
\highlightR{To enable the real-time evaluation of the MDCS-MPC algorithm at higher values of $\mu$, the control frequency is reduced to half of the switching frequency.
The triple-phase shift modulator then generates two consecutive sets of control angles for each output actuation of the controller.}

\begin{table}
    \vspace{-1em}
        \caption{Summary of experimental system parameters}\label{tab:parameters}
        \centering
        \begin{tabular}{c l l}\toprule
            Symbol & Description & Value \\
            \midrule
            \rowcolor{gray!25} $v_s$ & Input voltage from the dc source & \qty{300}{\volt} \\
            $C_1$ & Primary side dc capacitance & \qty{470}{\micro\farad} \\
            \rowcolor{gray!25} $C_2$ & Secondary side dc capacitance & \qty{470}{\micro\farad} \\
            $L$ & Inductance of the DAB & \qty{75.2}{\micro\henry} \\
            \rowcolor{gray!25} $n$ & Transformer turns ratio & 1.7 \\
            $f_{sw}$ & Switching frequency & \qty{100}{\kilo\hertz} \\
            \rowcolor{gray!25} $f_{c}$ & Control frequency & \qty{50}{\kilo\hertz} \\
            $\mu$ & MDCS-MPC discretization steps & 7 \\
            \rowcolor{gray!25}$\Delta_{max}$ & Maximum absolute voltage deviation & \qty{5}{\volt} \\
            $\gamma$ & Gain of the voltage-to-current actuation & 10 \\
            \rowcolor{gray!25}$\delta_i$ & Gain of the current actuation & 0.005 \\
            $\lambda$ & Objective function weighting factor & 7 \\
            \bottomrule
        \end{tabular}
    \end{table}

\begin{figure}[t]
    \centering
    \includegraphics[]{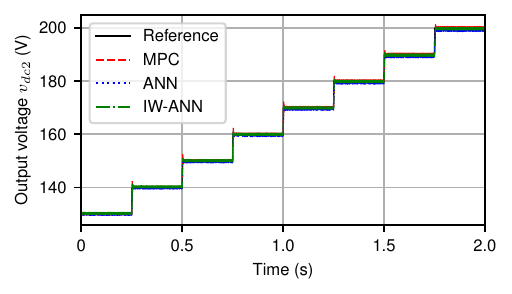}
    \vspace{-0.4cm}
    \caption{Steady-state output voltage tracking with MPC and ANN imitators trained with and without importance weighting.}\label{fig:experiment_output_voltage}
\end{figure}

The controllers were first tested with a $\qty{30.6}{\ohm}$ resistor as the output load of the DAB while tracking a sequence of output voltage reference steps.
The sequence was defined to last two seconds and run through a sequence of eight output voltage references between \qty{130}{\volt} and \qty{200}{\volt}, in \qty{10}{\volt} increments.
With the fast response time of the DAB converter, this test allows for the verification of both the steady-state performance and the transient response at different voltage levels.
Results are shown in Fig.~\ref{fig:experiment_output_voltage}, which displays the performance of three methods: the MDCS-MPC (MPC), and the neural network imitators trained with (ANN, weighted) and without (ANN) importance weighting.

\begin{figure}
    \centering
    \includegraphics[trim=6 0 0 0, clip]{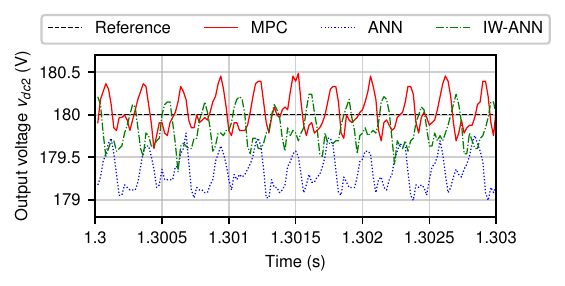}
    \vspace{-0.8cm}
    \caption{Zoomed-in section of Fig.~\ref{fig:experiment_output_voltage}, displaying the steady-state performance of the system. The conventionally-trained ANN model can be observed to result in a larger tracking error than the weighted model.}\label{fig:experiment_output_voltage_zoom}
\end{figure}

\begin{figure}
    \centering
    \includegraphics[]{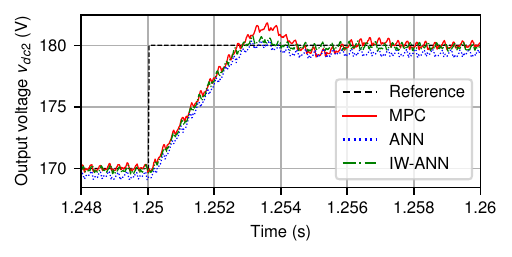}
    \vspace{-0.4cm}
    \caption{Zoomed-in section of Fig.~\ref{fig:experiment_output_voltage}, displaying the transient response to a voltage reference step between 170 V and 180 V.}\label{fig:experiment_output_voltage_transient}
\end{figure}

Fig.~\ref{fig:experiment_output_voltage_transient} shows a section of Fig.~\ref{fig:experiment_output_voltage} demonstrating the transient response of each method.
The MDCS-MPC presented a larger voltage overshoot than both ANN models, but was able to achieve a virtually zero low-frequency steady-state error.
During the transient itself, the two imitators resulted in a similar response, but Fig.~\ref{fig:experiment_output_voltage_transient} shows how the importance-weighted model was able to achieve a significantly reduced tracking error at the end of the transient when compared to the conventionally-trained model.
\highlightR{Despite these differences in performance, it must be noted that neither ANN model can be expected to outperform the MPC controller in a consistent manner.
As the models are trained to approximate the behavior of the MPC, they can at most perform as well as their imitation target.}

\begin{figure}
    \centering
    \includegraphics{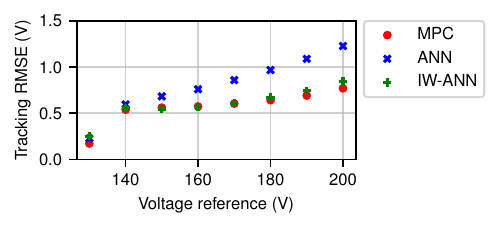}
    \vspace{-0.4cm}
    \highlightR{\caption{Voltage tracking RMSE at various output voltage levels for the considered controllers, based on the results shown in Fig. \ref{fig:experiment_output_voltage}.}
    \label{fig:voltage_error}}
\end{figure}

Although both the baseline model and the weighted model were able to track the reference voltage, they presented a small steady-state error when compared to the MPC algorithm, which is especially present at higher voltages.
The weighted model was able to significantly decrease this steady-state error when compared to the conventionally trained imitator.
This effect can be observed in Fig.~\ref{fig:experiment_output_voltage_zoom}, which shows a section of Fig.~\ref{fig:experiment_output_voltage} at a constant reference voltage.
\highlightR{Fig.~\ref{fig:voltage_error} quantifies the rms voltage tracking errors for each controller at the considered voltage level.
The figure shows how the baseline ANN model performs significantly worse in terms of steady-state error as the voltage increases, while the importance-weighted model is able to perform closer to the MPC.}

Table~\ref{tab:results} presents a final summary of these results, which further illustrate the effects of the importance-weighted training process.
By employing importance weights generated through the proposed density estimation method, the weighted model was able to reduce by a third the root mean square error (RMSE) of its predictions on $\mathcal{M}$, i.e. the experimental data set on which the KDE model was fit.
However, this came at the expense of a 10\% higher loss on average through the entirety of the input space, as defined by the uniformly sampled data set $\mathcal{D}$.
This observation confirms the theoretically described operation of the importance-weighted training process: the model was able to decrease its loss in the regions with a higher density of experimental data, at the cost of an increased average loss over the entire input space.   
Nonetheless, the net result was positive: the third row shows that the weighted model achieved a significantly lower experimental output voltage RMSE than the baseline imitator, bringing it much closer to the performance of the MPC algorithm.

\begin{table}
    \vspace{-1em}
        \caption{Controller comparison}\label{tab:results}
        \centering
        \begin{tabular}{l c c c}\toprule
            Metric & ANN & ANN, IW & MPC \\
            \midrule 
            \rowcolor{gray!25}$\bar{i}_2^{k}$ RMSE on $\mathcal{D}$ relative to MPC & $\mathbf{0.152}$ & $0.168$ & --\\ \addlinespace[0.1em]
            $\bar{i}_2^{k}$ RMSE on $\mathcal{M}$ relative to MPC & $0.0385$ & $\mathbf{0.0259}$ & --\\ \addlinespace[0.1em]
            \rowcolor{gray!25}Experimental ${v}_{dc2}$ tracking RMSE  & $0.853$ & $0.622$ & $\mathbf{0.594}$ \\ \addlinespace[0.1em]
            \bottomrule
        \end{tabular}
    \end{table}

\begin{figure}
    \centering
    \includegraphics[]{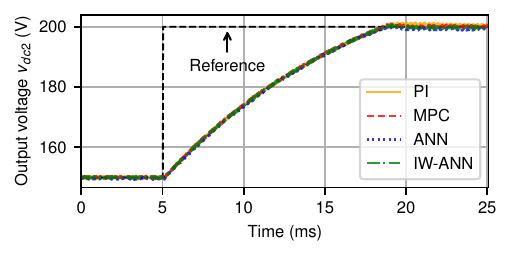}
    \vspace{-0.4cm}
    \highlightR{\caption{Response to a 50 V step reference.}
    \label{fig:experiment_output_voltage_transient_large}}
\end{figure}

\begin{figure}
    \centering
    \includegraphics[]{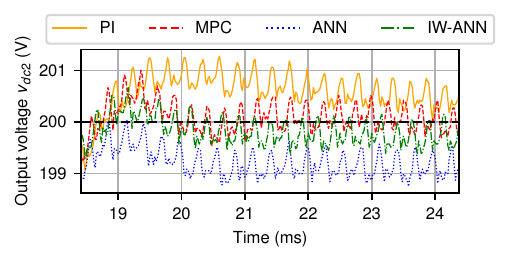}
    \vspace{-0.4cm}
    \highlightR{\caption{Zoomed-in section of Fig.~\ref{fig:experiment_output_voltage_transient_large}, displaying the settling of the output voltage with considered controller under analysis. The PI controller presents a larger settling time, while the conventionally-trained ANN model shows a steady-state error of approximately 1 V. }\label{fig:experiment_output_voltage_transient_large_zoom}}
\end{figure}

\highlightR{\subsubsection{PI control}
In order to evaluate the performance of the proposed method when compared to a more conventional control methodology, experimental results were collected using a PI controller for output voltage control.
The controller was tuned empirically using the laboratory setup to a proportional constant $K_p = 0.05$, and an integral time constant $T_i = \qty{200}{\micro s}$.
The PI controller employs the same triple-phase shift modulator as the other considered control algorithms.
Fig.~\ref{fig:experiment_output_voltage_transient_large} shows how the four controllers under consideration are able to similarly track a \qty{50}{V} reference step.
When zooming in on the overshoot of the output voltage, as shown in Fig.~\ref{fig:experiment_output_voltage_transient_large_zoom}, the MPC algorithm presents the smallest settling time, which is accurately approximated by the IW-ANN controller.}

\begin{figure}
    \centering
    \includegraphics[]{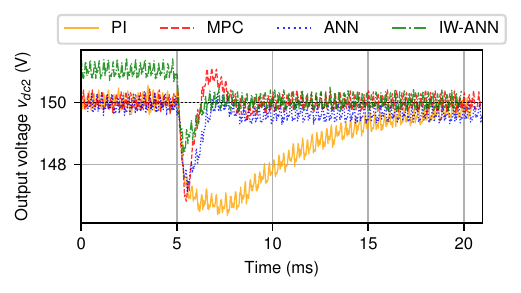}
    \vspace{-0.8cm}
    \highlightR{\caption{Resistive load impact from open-circuited output to $R = \qty{30.6}{\ohm}$ at $t =$ 5ms.}
    \label{fig:experiment_output_voltage_load_impact_off-to-on}}
\end{figure}

\highlightR{\subsubsection{Load impact}
In addition to voltage reference steps, the performance of the controllers was analyzed under load change conditions.
In the measurements shown in Fig.~\ref{fig:experiment_output_voltage_load_impact_off-to-on}, the \qty{30.6}{\ohm} resistive load was connected at the \qty{5}{ms} mark, from open-circuit conditions at the output.
Under these conditions, the PI controller can be observed to obtain a significantly slower voltage response than the other considered controllers, while MPC achieves a fast settling time with an overshoot of approximately 1 V.
The conventionally trained ANN presents a steady-state error after the load impact that is consistent with the previous findings.
Finally, the importance-weighted ANN shows fast dynamic response and accurate steady-state tracking under load; however, it presents a relatively large steady-state error before the load impact.
This behavior is to be expected, as the importance weights were obtained under the assumption of the presence of this resistive load.
Consequently, the model assigned a low importance to the open circuit output conditions, leading to worse performance in this region.}

\highlightR{\subsubsection{Parameter mismatch}
An important consideration when employing MPC algorithms is their sensitivity to model parameters.
Physical parameter estimation errors will typically result in worse control performance, as the controller may consistently overestimate or underestimate state variables, leading to suboptimal control actions.
Initial errors can also be amplified by component aging, which will often significantly affect the behavior of passive components.

With the aim to study the effects of parameter mismatch on the considered control strategies, multiple tests were carried out with a modified experimental setup, with the value of the high-frequency inductance $L = \qty{75.2}{\micro\henry}$ reduced by about 20\% to $L = \qty{60.5}{\micro\henry}$.
Figs. \ref{fig:experiment_output_voltage_transient_mismatch_MPC}, \ref{fig:experiment_output_voltage_transient_mismatch_ANN}, and \ref{fig:experiment_output_voltage_transient_mismatch_IW_ANN} show the response to a voltage reference step with the MDCS-MPC controller, the conventionally-trained ANN imitator, and the importance-weighted ANN, respectively. 
For these experiments, the value of the inductance in the MPC algorithm was kept at $L = \qty{75.2}{\micro\henry}$, and the two ANN controllers were trained to imitate this same MPC strategy.

\begin{figure}
    \centering
    \includegraphics[]{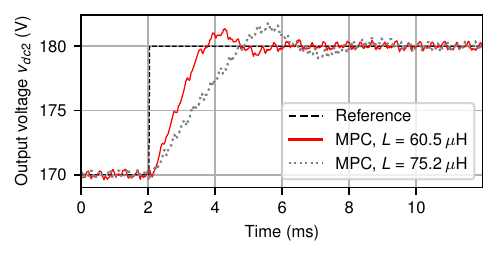}
    \vspace{-0.4cm}
    \highlightR{\caption{Transient response with MPC under parameter mismatch.}
    \label{fig:experiment_output_voltage_transient_mismatch_MPC}}
\end{figure}

\begin{figure}
    \centering
    \includegraphics[]{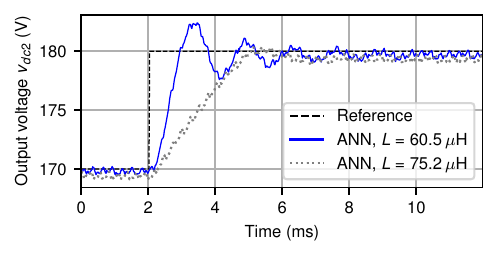}
    \vspace{-0.4cm}
    \highlightR{\caption{Transient response with the conventionally-trained ANN under parameter mismatch.}
    \label{fig:experiment_output_voltage_transient_mismatch_ANN}}
\end{figure}

\begin{figure}
    \centering
    \includegraphics[]{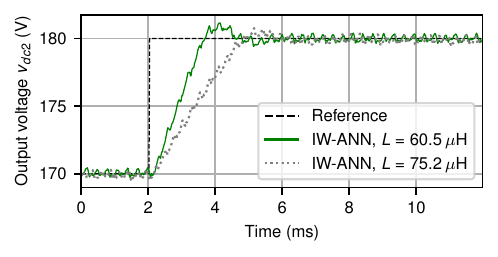}
    \vspace{-0.4cm}
    \highlightR{\caption{Transient response with the importance-weighted ANN under parameter mismatch.}
    \label{fig:experiment_output_voltage_transient_mismatch_IW_ANN}}
\end{figure}

Under these conditions, the MPC controller (in Fig.~\ref{fig:experiment_output_voltage_transient_mismatch_MPC}) and the IW-ANN (in Fig.~\ref{fig:experiment_output_voltage_transient_mismatch_IW_ANN}) are both able to retain adequate performance despite the inaccuracy of the model, with a faster response due to the decreased inductance and negligible differences in settling time and steady-state error.
While the IW-ANN shows a slightly higher overshoot, this behavior is much more pronounced with the conventionally-trained ANN (in Fig.~\ref{fig:experiment_output_voltage_transient_mismatch_ANN}), which additionally presents a more significant oscillatory behavior and retains its steady-state error.
Under the considered conditions, these results suggest that the MPC algorithm is robust to parameter mismatch in its system model.
This property is also found in the IW-ANN model, which approximates the MPC algorithm more closely in the considered operating points than the conventionally-trained ANN, which appears to be less resilient to parameter mismatch.
}

\subsection{Computational effort}
As imitator controllers can only be expected to approximate the behavior of the target MPC algorithm, their main advantage is the potential reduction of computational requirements, which enables real-time deployment in resource-constrained computing platforms.
The model size of the ANN-based imitators then becomes a key consideration: larger models can achieve a higher prediction accuracy and therefore better approximate the target algorithm, but their evaluation is more computationally demanding, both in terms of memory and number of operations.
The proposed training method can serve as a tool to balance this trade-off, by allowing for smaller ANNs to achieve comparable performance to larger models in the targeted regions of their input space.

\begin{figure}
    \centering
    \includegraphics[]{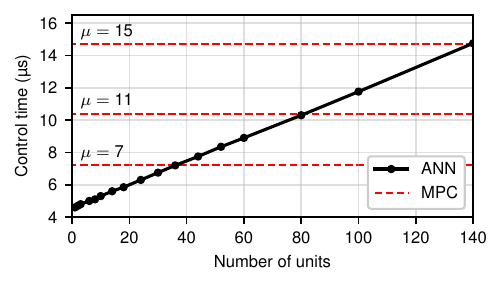}
    \vspace{-0.4cm}
    \caption{Time requirements for the execution of a single step of the control algorithms on the Imperix B-Box RCP platform. \highlightR{The measured execution times remained practically constant across multiple iterations of each algorithms: the graph therefore presents individual samples.} The horizontal axis represents the number of units in the hidden layer of the ANN imitators, while each dashed line corresponds to the MDCS-MPC at fixed discretization steps $\mu$.}\label{fig:control_time}
\end{figure}

To compare the computational requirements of the ANN-based controllers and the MDCS-MPC, the execution time of the two control algorithms was measured on the Imperix B-Box RCP platform, which uses an AMD Zynq XC7030-3 system-on-chip (SoC).
The resulting measurements are graphed in Fig.~\ref{fig:control_time}.
The execution time of the ANN-based controller was obtained for different sizes of the single hidden layer, given by its number of units.
The control time of the MDCS-MPC is proportional to both its prediction horizon and its discretization step $\mu$: for this comparison, the horizon was fixed at two-step predictions and measurements were recorded at several values of $\mu$.
In the voltage tracking tests, where the ANN controllers employed 10 hidden units, the average control time was \qty{5.3}{\micro\s}, a ${\sim}26\%$ reduction from the \qty{7.2}{\micro\s} required by the MDCS-MPC with $\mu = 7$.
As the training data can be generated offline and model size can be fixed, the ANN controllers can mimic MDCS-MPC with larger values of $\mu$, even if the resulting MPC implementation becomes unable to run in real-time.  

\section{Conclusion}\label{sec:conclusion}
The present article proposes a method to improve model performance in the growing framework of MPC imitation learning.
This enhancement is achieved through importance-weighted training, which allows designers to target specific regions of the input space during training, thus making better use of the commonly limited capacity of the MPC approximator models.
Importance weights penalize the training loss in certain regions of the input space more than in others, leading to higher average loss over the entirety of the input space but to lower loss in the targeted regions.
Adequately designing the importance weights is therefore crucial and one of the key points of the proposed method, which obtains their values through kernel density estimation.

In the conventional supervised learning approach for MPC approximation, the designer either has little control over how the model learns in the input space (in the case of sampling methods) or too much responsibility in designing simulations or experiments that can cover the desired regions while allowing for a certain degree of generalization.
The proposed method bridge this gap by combining the comprehensive coverage of the input space provided by sampling methods with the application-dependent information about which regions are most relevant contained in simulation or experimental data.
The presented implementation introduces only a single additional tunable parameter, the bandwidth of the kernel, allowing for control over the generalization-exploitation trade-off with limited additional complexity.
The proposed method is especially relevant for high-frequency or low-compute applications, where MPC imitators must be designed to be as computationally efficient as possible: the method can then serve to maintain performance in the main regions of operation of the converter.
\highlightR{Despite its potential improvements over conventional imitator training, the proposed method introduces additional complexity to the training process and requires careful consideration of the input space regions to be targeted.
As observed experimentally, although the importance-weighted imitator controller performs better under the selected conditions, it will typically underperform outside of them when compared to a conventionally trained model: it is therefore crucial to thoroughly design and test the trained controllers both inside and outside the regions of interest. 
In addition, verifying the stability of ANN-based controllers is also an active area of research which is key for the widespread adoption of these methods in industrial applications.
Future work could focus on analyzing the system stability of the converter inside its safe operating area when using this type of controllers.}

Future developments of the proposed method may include its application to different types of MPC, such as classification problems in the form of finite control set MPC, verification of the robustness of importance-weighted models in a variety of operating conditions, or the use of alternative importance weight definitions aiming to achieve additional goals.
\highlightR{A robust method to tune the kernel bandwidth $h$ with minimal reliance on empirical observations would also be an important specific improvement to the proposed method.}
\highlightR{Additionally, while the present article is limited in scope to a DAB converter in stand-alone operation, future research should consider the interactions between multiple converters within more complex systems.}

\bibliographystyle{Bibliography/IEEEtranTIE}
\bibliography{Bibliography/IEEEabrv,Bibliography/BIB_24-TIE-0908}\

\begin{thebibliography}{10}
\providecommand{\url}[1]{#1}
\csname url@samestyle\endcsname
\providecommand{\newblock}{\relax}
\providecommand{\bibinfo}[2]{#2}
\providecommand{\BIBentrySTDinterwordspacing}{\spaceskip=0pt\relax}
\providecommand{\BIBentryALTinterwordstretchfactor}{4}
\providecommand{\BIBentryALTinterwordspacing}{\spaceskip=\fontdimen2\font plus
\BIBentryALTinterwordstretchfactor\fontdimen3\font minus \fontdimen4\font\relax}
\providecommand{\BIBforeignlanguage}[2]{{%
\expandafter\ifx\csname l@#1\endcsname\relax
\typeout{** WARNING: IEEEtran.bst: No hyphenation pattern has been}%
\typeout{** loaded for the language `#1'. Using the pattern for}%
\typeout{** the default language instead.}%
\else
\language=\csname l@#1\endcsname
\fi
#2}}
\providecommand{\BIBdecl}{\relax}
\BIBdecl

\bibitem{Vazquez2014}
S.~Vazquez, J.~I. Leon, L.~G. Franquelo, J.~Rodriguez, H.~A. Young, A.~Marquez, and P.~Zanchetta, ``Model predictive control: A review of its applications in power electronics,'' \emph{IEEE Industrial Electronics Magazine}, vol.~8, \href{http://dx.doi.org/10.1109/MIE.2013.2290138}{DOI 10.1109/MIE.2013.2290138}, pp. 16--31, 2014.

\bibitem{Vazquez2017}
S.~Vazquez, J.~Rodriguez, M.~Rivera, L.~G. Franquelo, and M.~Norambuena, ``Model predictive control for power converters and drives: Advances and trends,'' \emph{IEEE Transactions on Industrial Electronics}, vol.~64, \href{http://dx.doi.org/10.1109/TIE.2016.2625238}{DOI 10.1109/TIE.2016.2625238}, pp. 935--947, 2 2017.

\bibitem{Novak2021}
M.~Novak and T.~Dragicevic, ``Supervised imitation learning of finite-set model predictive control systems for power electronics,'' \emph{IEEE Transactions on Industrial Electronics}, vol.~68, \href{http://dx.doi.org/10.1109/TIE.2020.2969116}{DOI 10.1109/TIE.2020.2969116}, pp. 1717--1723, 2 2021.

\bibitem{Wang2021a}
S.~Wang, T.~Dragicevic, G.~F. Gontijo, S.~K. Chaudhary, and R.~Teodorescu, ``Machine learning emulation of model predictive control for modular multilevel converters,'' \emph{IEEE Transactions on Industrial Electronics}, vol.~68, \href{http://dx.doi.org/10.1109/TIE.2020.3038064}{DOI 10.1109/TIE.2020.3038064}, pp. 11\,628--11\,634, 11 2021.

\bibitem{Wang2021b}
S.~Wang, T.~Dragicevic, Y.~Gao, and R.~Teodorescu, ``Neural network based model predictive controllers for modular multilevel converters,'' \emph{IEEE Transactions on Energy Conversion}, vol.~36, \href{http://dx.doi.org/10.1109/TEC.2020.3021022}{DOI 10.1109/TEC.2020.3021022}, pp. 1562--1571, 6 2021.

\bibitem{Zafra2024}
E.~Zafra, J.~Granado, V.~B. Lecuyer, S.~Vazquez, A.~M. Alcaide, J.~I. Leon, and L.~G. Franquelo, ``Computationally efficient sphere decoding algorithm based on artificial neural networks for long-horizon fcs-mpc,'' \emph{IEEE Transactions on Industrial Electronics}, vol.~71, \href{http://dx.doi.org/10.1109/TIE.2023.3243301}{DOI 10.1109/TIE.2023.3243301}, pp. 39--48, 1 2024.

\bibitem{Simonetti2023}
F.~Simonetti, A.~D'Innocenzo, and C.~Cecati, ``Neural network model-predictive control for chb converters with fpga implementation,'' \emph{IEEE Transactions on Industrial Informatics}, vol.~19, \href{http://dx.doi.org/10.1109/TII.2023.3233973}{DOI 10.1109/TII.2023.3233973}, pp. 9691--9702, 9 2023.

\bibitem{Mohamed2019}
I.~S. Mohamed, S.~Rovetta, T.~D. Do, T.~Dragicevic, and A.~A. Diab, ``A neural-network-based model predictive control of three-phase inverter with an output {LC} filter,'' \emph{IEEE Access}, vol.~7, \href{http://dx.doi.org/10.1109/ACCESS.2019.2938220}{DOI 10.1109/ACCESS.2019.2938220}, pp. 124\,737--124\,749, 2019.

\bibitem{Wang2022}
D.~Wang, Z.~J. Shen, X.~Yin, S.~Tang, X.~Liu, C.~Zhang, J.~Wang, J.~Rodriguez, and M.~Norambuena, ``Model predictive control using artificial neural network for power converters,'' \emph{IEEE Transactions on Industrial Electronics}, vol.~69, \href{http://dx.doi.org/10.1109/TIE.2021.3076721}{DOI 10.1109/TIE.2021.3076721}, pp. 3689--3699, 4 2022.

\bibitem{Liu2023}
X.~Liu, L.~Qiu, Y.~Fang, and J.~Rodriguez, ``Predictor-based data-driven model-free adaptive predictive control of power converters using machine learning,'' \emph{IEEE Transactions on Industrial Electronics}, vol.~70, \href{http://dx.doi.org/10.1109/TIE.2022.3208594}{DOI 10.1109/TIE.2022.3208594}, pp. 7591--7603, 8 2023.

\bibitem{Wang2020}
Z.~Wang, H.~X. Li, and C.~Chen, ``Incremental reinforcement learning in continuous spaces via policy relaxation and importance weighting,'' \emph{IEEE Transactions on Neural Networks and Learning Systems}, vol.~31, \href{http://dx.doi.org/10.1109/TNNLS.2019.2927320}{DOI 10.1109/TNNLS.2019.2927320}, pp. 1870--1883, 6 2020.

\bibitem{Fedeli2023}
F.~Fedeli, A.~M. Metelli, F.~Trovo, and M.~Restelli, ``{IWDA}: Importance weighting for drift adaptation in streaming supervised learning problems,'' \emph{IEEE Transactions on Neural Networks and Learning Systems}, vol.~34, \href{http://dx.doi.org/10.1109/TNNLS.2023.3265524}{DOI 10.1109/TNNLS.2023.3265524}, pp. 6813--6823, 10 2023.

\bibitem{Chen2020}
L.~Chen, L.~Lin, S.~Shao, F.~Gao, Z.~Wang, P.~W. Wheeler, and T.~Dragicevic, ``Moving discretized control set model-predictive control for dual-active bridge with the triple-phase shift,'' \emph{IEEE Transactions on Power Electronics}, vol.~35, \href{http://dx.doi.org/10.1109/TPEL.2019.2962838}{DOI 10.1109/TPEL.2019.2962838}, pp. 8624--8637, 8 2020.

\bibitem{Chen2020a}
L.~Chen, S.~Shao, Q.~Xiao, L.~Tarisciotti, P.~W. Wheeler, and T.~Dragicevic, ``Model predictive control for dual-active-bridge converters supplying pulsed power loads in naval dc micro-grids,'' \emph{IEEE Transactions on Power Electronics}, vol.~35, \href{http://dx.doi.org/10.1109/TPEL.2019.2917450}{DOI 10.1109/TPEL.2019.2917450}, pp. 1957--1966, 2 2020.

\bibitem{Chen2020b}
L.~Chen, F.~Gao, K.~Shen, Z.~Wang, L.~Tarisciotti, P.~Wheeler, and T.~Dragicevic, ``Predictive control based dc microgrid stabilization with the dual active bridge converter,'' \emph{IEEE Transactions on Industrial Electronics}, vol.~67, \href{http://dx.doi.org/10.1109/TIE.2020.2965460}{DOI 10.1109/TIE.2020.2965460}, pp. 8944--8956, 10 2020.

\bibitem{Lopez2023}
M.~L\'{o}pez, N.~Mijatovic, J.~Rodr\'{i}guez, and T.~Dragi\v{c}evi\'{c}, ``Model predictive voltage and current control of dual active bridge using enhanced moving discretized control set,'' \emph{IECON 2023- 49th Annual Conference of the IEEE Industrial Electronics Society}, \href{http://dx.doi.org/10.1109/IECON51785.2023.10312249}{DOI 10.1109/IECON51785.2023.10312249}, pp. 1--8, 10 2023.

\bibitem{Shao2019}
S.~Shao, M.~Jiang, W.~Ye, Y.~Li, J.~Zhang, and K.~Sheng, ``Optimal phase-shift control to minimize reactive power for a dual active bridge dc-dc converter,'' \emph{IEEE Transactions on Power Electronics}, vol.~34, \href{http://dx.doi.org/10.1109/TPEL.2018.2890292}{DOI 10.1109/TPEL.2018.2890292}, pp. 10\,193--10\,205, 10 2019.

\bibitem{Cortes2012}
P.~Cortes, J.~Rodriguez, C.~Silva, and A.~Flores, ``Delay compensation in model predictive current control of a three-phase inverter,'' \emph{IEEE Transactions on Industrial Electronics}, vol.~59, \href{http://dx.doi.org/10.1109/TIE.2011.2157284}{DOI 10.1109/TIE.2011.2157284}, pp. 1323--1325, 2 2012.

\bibitem{Goodfellow2016}
I.~Goodfellow, Y.~Bengio, and A.~Courville, \emph{Deep Learning}.\hskip 1em plus 0.5em minus 0.4em\relax MIT Press, 2016.

\bibitem{Fang2020}
T.~Fang, N.~Lu, G.~Niu, and M.~Sugiyama, ``Rethinking importance weighting for deep learning under distribution shift,'' \emph{Advances in Neural Information Processing Systems}, vol. 2020-December, 6 2020.

\bibitem{Kingma2014}
D.~P. Kingma and J.~L. Ba, ``Adam: A method for stochastic optimization,'' \emph{3rd International Conference on Learning Representations, ICLR 2015 - Conference Track Proceedings}, \href{http://dx.doi.org/10.48550/arxiv.1412.6980}{DOI 10.48550/arxiv.1412.6980}, 12 2015.

\bibitem{Byrd2018}
J.~Byrd and Z.~C. Lipton, ``What is the effect of importance weighting in deep learning?'' \emph{36th International Conference on Machine Learning, ICML 2019}, vol. 2019-June, \href{http://dx.doi.org/10.48550/arxiv.1812.03372}{DOI 10.48550/arxiv.1812.03372}, pp. 1405--1419, 12 2018.

\bibitem{Xu2021}
D.~Xu, Y.~Ye, and C.~Ruan, ``Understanding the role of importance weighting for deep learning,'' \emph{ICLR 2021 - 9th International Conference on Learning Representations}, \href{http://dx.doi.org/10.48550/arxiv.2103.15209}{DOI 10.48550/arxiv.2103.15209}, 3 2021.

\bibitem{Silverman2018}
B.~Silverman, \emph{Density Estimation for Statistics and Data Analysis}.\hskip 1em plus 0.5em minus 0.4em\relax Routledge, 2 2018.

\bibitem{He2015}
K.~He, X.~Zhang, S.~Ren, and J.~Sun, ``Delving deep into rectifiers: Surpassing human-level performance on imagenet classification,'' \emph{2015 IEEE International Conference on Computer Vision (ICCV)}, \href{http://dx.doi.org/10.1109/ICCV.2015.123}{DOI 10.1109/ICCV.2015.123}, pp. 1026--1034, 12 2015.

\bibitem{McKay1979}
M.~D. McKay, R.~J. Beckman, and W.~J. Conover, ``A comparison of three methods for selecting values of input variables in the analysis of output from a computer code,'' \emph{Technometrics}, vol.~21, \href{http://dx.doi.org/10.2307/1268522}{DOI 10.2307/1268522}, p. 239, 5 1979.

\bibitem{scikit-learn}
F.~Pedregosa, G.~Varoquaux, A.~Gramfort, V.~Michel, B.~Thirion, O.~Grisel, M.~Blondel, P.~Prettenhofer, R.~Weiss, V.~Dubourg, J.~Vanderplas, A.~Passos, D.~Cournapeau, M.~Brucher, M.~Perrot, and E.~Duchesnay, ``Scikit-learn: Machine learning in {P}ython,'' \emph{Journal of Machine Learning Research}, vol.~12, pp. 2825--2830, 2011.

\end{thebibliography}

\begin{IEEEbiography}[{\includegraphics[width=1in,height=1.25in,clip,keepaspectratio]{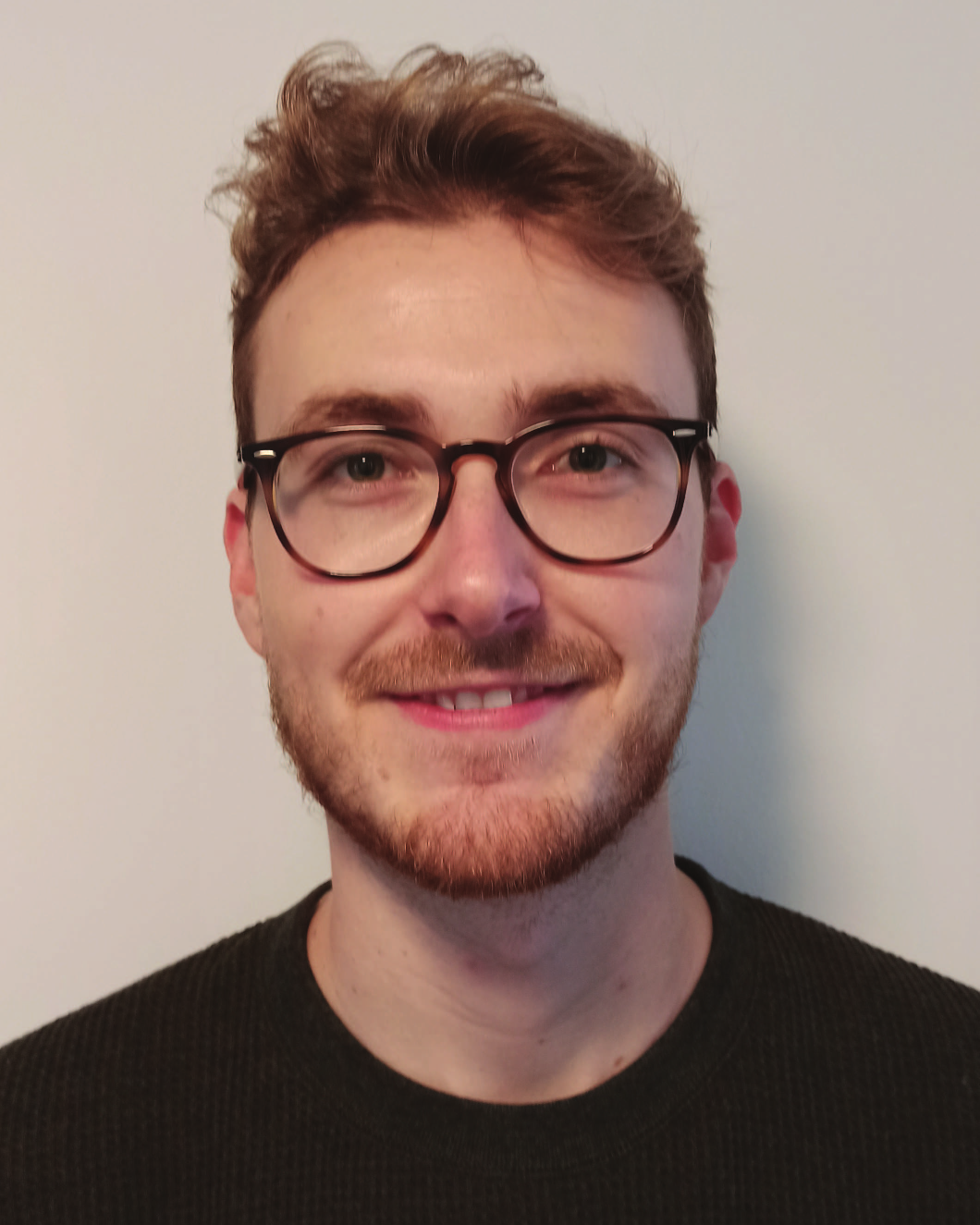}}]
{Pere Izquierdo G\'omez} (Student Member, IEEE) received the B.Sc.\ degree in energy engineering from the Polytechnic University of Catalonia, Barcelona, Spain, in 2018, the M.Sc.\ degree in energy engineering from Aalborg University, Aalborg, Denmark, in 2020, and the Ph.D. degree in electrical engineering from the Technical University of Denmark, Kongens Lyngby, Denmark, in 2024.
He is currently a Postdoctoral Researcher in the Department of Wind and Energy Systems of the Technical University of Denmark, with a focus on machine learning applications to power converter control and diagnostics.
\end{IEEEbiography}

\vspace{-0.5cm}
\begin{IEEEbiography}[{\includegraphics[width=1in,height=1.25in,clip,keepaspectratio]{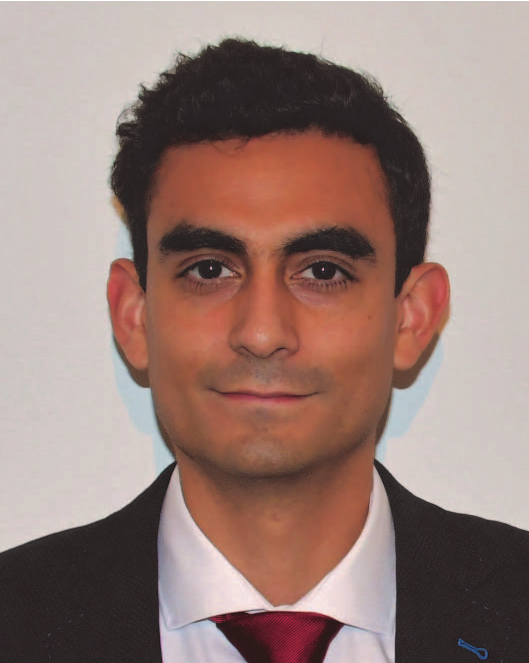}}]
{Miguel E. L\'opez Gajardo} (Student Member, IEEE) received the B.Sc.\ (Ingeniero Civil Electr\'onico) and the M.Sc.\ degrees in electronic engineering from the Universidad T\'ecnica Federico Santa Mar\'ia, Valpara\'iso, Chile, in 2014. He is currently working toward the Ph.D.\ degree in electrical engineering with the Department of Wind and Energy Systems, Technical University of Denmark, Kongens Lyngby, Denmark. From 2015 to 2020, he was an R\&D Engineer with the Advanced Center for Electrical and Electronic Engineering. His research interests include power electronics, e-mobility, renewable energy conversion systems, drives, AI, and their applications.
\end{IEEEbiography}

\vspace{3.4cm}
\begin{IEEEbiography}[{\includegraphics[width=1in,height=1.25in,clip,keepaspectratio]{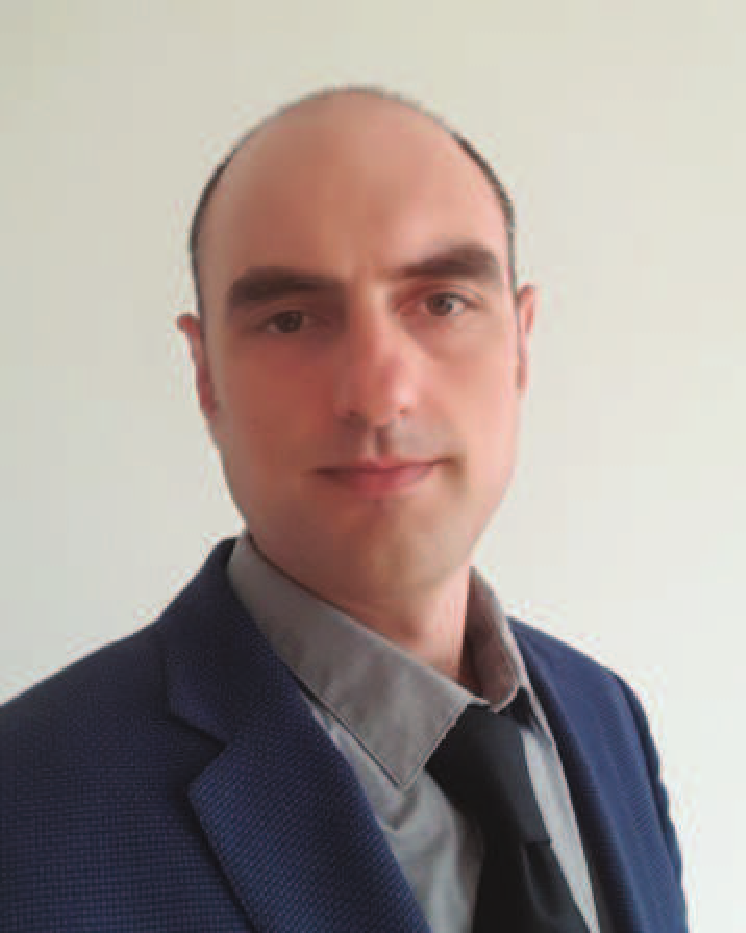}}]
{Nenad Mijatovic} (Senior Member, IEEE) received the Dipl.Ing. degree in electrical power engineering from the University of Belgrade, Belgrade, Serbia, in 2007, and the Ph.D.\ degree in electrical engineering from the Technical University of Denmark, Kongens Lyngby, Denmark, in 2012, for his work on the technical feasibility of novel machines and drives for the wind industry.
He was an Industrial Postdoc in the field of wind turbine direct-drive concepts. He is currently an Associate Professor with the Technical University of Denmark, where he is in charge of managing research projects and education related to the field of electrical machines and drives, power electronic converters, motion control, application of energy storage and general applications of low-frequency electromagnetism, large-scale application of superconductivity, with a main focus on emerging eMobility and renewable energy generation, novel electrical machine drives/actuator designs, operation, control and diagnostics of electromagnetic assemblies, advanced control of drives and grid-connected power electronics, energy storage, and eMobility. Dr.\ Mijatovic has been a Member of IEEE since 2008 and a Senior Member of IEEE since 2018. 
\end{IEEEbiography}

\vspace{-0.5cm}
\begin{IEEEbiography}[{\includegraphics[width=1in,height=1.25in,clip,keepaspectratio]{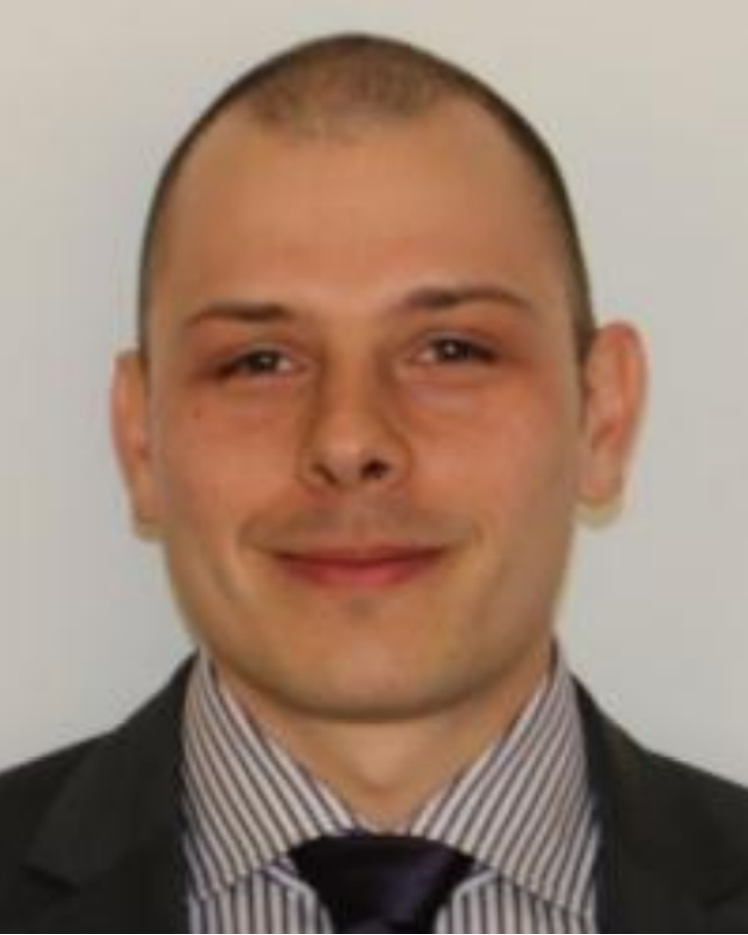}}]
{Tomislav Dragi\v{c}evi\'{c}} (Senior Member, IEEE) received the M.Sc.\ and industrial Ph.D.\ degrees in electrical engineering from the Faculty of Electrical Engineering, Zagreb, Croatia, in 2009 and 2013, respectively. He is currently a Professor with the Technical University of Denmark, Kongens Lyngby, Denmark, where he leads research in the digitalization of power converters. He has authored or coauthored more than 150 IEEE journal articles in these areas. Dr.\ Dragi\v{c}evi\'{c} is currently an associate editor of \textsc{IEEE Transactions on Industrial Electronics} and \textsc{IEEE Emerging and Selected Topics in Power Electronics} and an Editor of \textit{IEEE Industrial Electronics Magazine}. He is a Senior Member of the IEEE.
\end{IEEEbiography}

\end{document}